# A Mathematical Framework for Estimating Risk of Airborne Transmission of COVID-19 with Application to Facemask Use and Social Distancing

[1,2]Rajat Mittal (रजत मित्तल), [1]Charles Meneveau and [3]Wen Wu (武文)

[1]*Mechanical Engineering, Johns Hopkins University, 3400 N. Charles St., Baltimore, Maryland, 21218*
[2]*School of Medicine, Johns Hopkins University, 3400 N. Charles St., Baltimore, Maryland, 21218*
[3]*Mechanical Engineering, University of Mississippi, 201C Carrier Hall, Oxford, Mississippi, 38677*

## 1 Abstract

A mathematical model for estimating the risk of airborne transmission of a respiratory infection such as COVID-19, is presented. The model employs basic concepts from fluid dynamics and incorporates the known scope of factors involved in the airborne transmission of such diseases. Simplicity in the mathematical form of the model is by design, so that it can serve not only as a common basis for scientific inquiry across disciplinary boundaries, but also be understandable by a broad audience outside science and academia. The caveats and limitations of the model are discussed in detail. The model is used to assess the protection from transmission afforded by face coverings made from a variety of fabrics. The reduction in transmission risk associated with increased physical distance between the host and susceptible is also quantified by coupling the model with available data on scalar dispersion in canonical flows. Finally, the effect of the level of physical activity (or exercise intensity) of the host and the susceptible in enhancing transmission risk, is also assessed.

## 2 Introduction

COVID-19 spread across the world with a speed and intensity that laid bare the limits of our understanding of the transmission pathways and the associated factors that are key to the spread of such diseases. There is however an emerging consensus that "airborne transmission," where virion bearing respiratory droplets and droplet nuclei (also called respiratory aerosols) expelled by an infected person (the ``host'') are inhaled by a "susceptible" individual, constitutes an important mode for the spread of COVID-19[1–5]. Questions regarding the size of the droplets involved[6–9] and the range of such transmission[10] can be bypassed by noting that the key element that differentiates airborne transmission from the droplet and contact routes of transmission[11] is the essential role of *inhalation* by the susceptible in this pathway for transmission. Generally, it is the small (<10 μm) particles that are likely to be entrained into the inhalation current of a person, but environmental conditions as well as the proximity between the host and the susceptible could allow larger particles/droplets to play a role in airborne transmission.

Irrespective of the size of droplets or the range involved, airborne transmission of COVID-19 and other respiratory infections involve the following sequence of events (see Fig. 1):

1. generation, expulsion and aerosolization of virus-containing droplets from the mouth and nose of an infected host;
2. dispersion and transport via ambient air currents of this respiratory aerosol to a susceptible; and
3. inhalation of droplets/aerosols, and deposition of virus in the respiratory mucosa of the susceptible.

Each phase in this sequence has complex dependencies on a variety of factors that may include the morphological properties and pathogenicity of the virus, the health status of the host and/or the susceptible, environmental conditions, and the presence/effectiveness of face coverings being used by the host and/or susceptible. Given this complexity of phenomenology and the many factors involved, it is not surprising that even after more than 8 months of the world dealing with the COVID-19 pandemic, there are fundamental questions that continue to confound scientists, policy makers and the members of the public at-large. These include questions such as: what factors have enabled the SARS-CoV-2 to spread so much faster and more extensively than other similar viruses in the recent past[12,13]? Why is the rate of infection so different in different regions/countries of the world[14]? How much lower is the likelihood of transmission in an outdoor environment compared to an indoor environment[10,15]? How do policies and societal behavior such as compliance with mask wearing affect the rate of transmission[16,17]? And finally, how does transmission risk reduce with distance between the host and the susceptible?

---

[1] mittal@jhu.edu

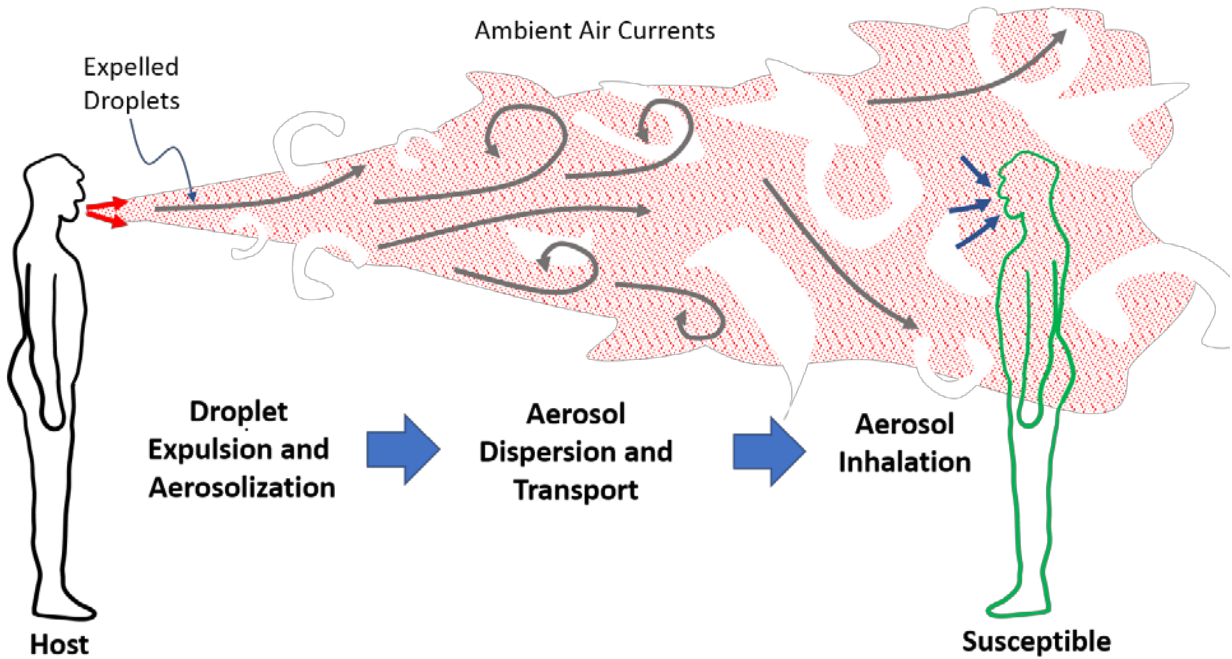

*Figure 1. Schematic depicting the key stages in the airborne transmission of a respiratory infection such as COVID-19.*

Scientists spanning fields such as biomedicine, epidemiology, virology, public health, fluid dynamics, aerosol physics, public policy, behavioral psychology, and others, are tackling these as well as other important questions. However, what is lacking is a simple and intuitive conceptual framework (or model) that encapsulates the complex, multifactorial scope of this problem in a manner that not only serves as a common basis for scientific inquiry across disciplinary boundaries, but also as a tool to more easily communicate the factors associated with the spread of this disease, to a wide range of stakeholders including non-scientists such as policy-makers, public media, and the public at-large. Given the rapidly evolving nature of the pandemic and the resurgence of infections in many commmunities[18], the importance of clear communication of infection risk across scientific disciplines, as well as to policy/decision makers and other segments of society, is more important than ever.

## 3    The Contagion Airborne Transmission (CAT) Inequality

In 1961, Dr. Frank Drake, an astronomer and astrophysicist involved in the search for extraterrestrial intelligence, conceived a conceptual framework to predict the number of technological civilizations that may exist in our galaxy. The Drake Equation[19,20], as it has become known, involves a number of probabilistic factors, which when multiplied together, result in the number of civilizations within our galaxy, at any given moment, that humanity could communicate with. The power of this equation is not in that it actually allows us to predict this number with a known level of certainty, but in the fact that it provides an easy to understand framework for grasping the key factors involved in something that seems inestimable: the number of advanced lifeforms that exist elsewhere in our galaxy.

Motivated by the Drake Equation, and based on the idea that airborne transmission occurs if a susceptible inhales a viral dose that exceeds the minimum infectious dose[21,22], the following model to predict the possibility of airborne transmission of a respiratory contagion such as SARS-CoV-2 from an infected host to a susceptible, is proposed:

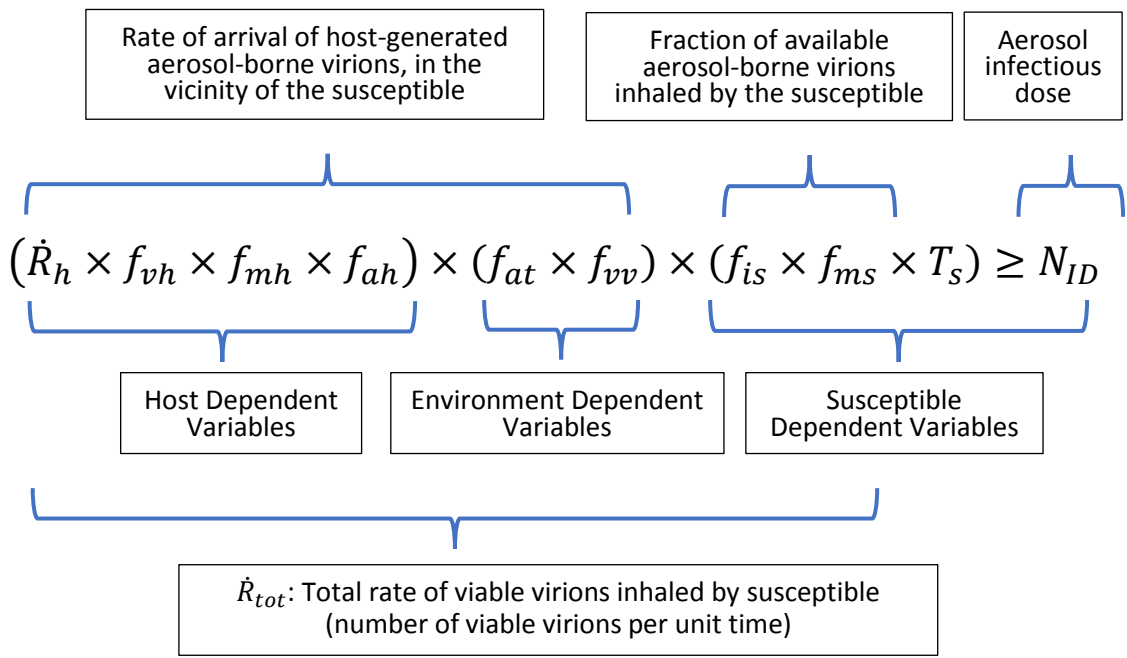

*Figure 2 – The contagion airborne transmission (CAT) inequality that evaluates the conditions for airborne transmission of a respiratory infection such as COVID-19. The left-hand-side of the inequality represents the total inhaled viral dose and the right hand side is the minimum aerosol dose required to initiate an infection in the susceptible. The inequality is satisfied (and transmission is successful) when the susceptible inhales a viral dose that exceeds the minimum infectious dose. The variables in the model can be segregated in different ways as shown in the graphic above.*

In the above expression:

$\dot{R}_h$: rate of expulsion of respiratory droplets from the nose and mouth of the host (number of droplets per unit time).

$f_{vh}$: fractional viral emission load – average number of virions contained in each expelled droplets

$f_{mh}$: fraction of expelled droplets that make it past the face-covering of the host

$f_{ah}$: fraction of expelled droplets that aerosolize (i.e. become suspended in the air)

$f_{at}$: fraction of aerosolized droplets that transport to the vicinity of the susceptible

$f_{vv}$: fraction of aerosolized droplets transported to the vicinity of the susceptible that contain viable virions

$f_{is}$: fraction of aerosols in the vicinity of the susceptible that would be inhaled by a susceptible not wearing a face covering

$f_{ms}$: fraction of inhaled aerosols that are filtered by the face covering of the susceptible.

$\dot{R}_{tot}$: total rate of viable virion inhalation by the susceptible (number per unit time).

$T_s$: duration of exposure of the susceptible to the aerosols from the host.

$N_{ID}$: minimum number of inhaled virions required to initiate infection in the susceptible.

The CAT inequality is a mathematical model for the estimation of infection risk that can be used based on various interpretations of its constituent terms. For example, if $\dot{R}_{tot}$ and $N_{ID}$ are known, then the CAT inequality allows one to deduce the critical exposure time $T_{sC} = N_{ID}/\dot{R}_{tot}$ below which infection is unlikely. Most often in practice, however, not all of the factors in the inequality will be known. Still, the CAT inequality can be used to compare *relative* risks since one may consider the risk of a situation to be inversely proportional to the corresponding critical exposure time (e.g. halving critical exposure time doubles the risk). For example, the risk ratio for two situations *A* and *B* will be $T_{sc-B}/T_{sc-A} = \dot{R}_{tot-A}/\dot{R}_{tot-B}$. Thus, if e.g. all factors are equal for *A* and *B* except for one, e.g. $f_{mhA} \neq f_{mhB}$ (say), then the risk ratio comparing *A* to *B* will be $f_{mhA}/f_{mhB}$. In Sec. 8 we will illustrate such relative risk estimations for three specific cases.

The use of mathematical models to predict infection rates is well-established in epidemiology[21,23,24] and the CAT inequality belongs among such models (see Sec. 8 for discussion on relationship to existing modeling frameworks). As with any model, the CAT inequality has a number of underlying assumptions (see Sec. 9) but its potential advantage is that it presents transmission risk via a simple mathematical expression that on one hand, captures a wide scope of factors that may be involved in airborne transmission, and on the other, is easy to convey to scientists from a wide range of fields, non-scientists such as policy makers, public officials, and public media, as well as even members of the general public.

As laid out in a previous publication,[11] each stage in the airborne transmission process is mediated by complex flow phenomena, ranging from air-mucous interaction and liquid sheet fragmentation inside the respiratory tract, to turbulence in the expiratory jet/ambient flow and flow-induced droplet evaporation and particle dispersion, to inhalation and deposition of aerosols in the lungs. Furthermore, non-pharmaceutical approaches employed to mitigate respiratory infections such as social distancing and the wearing of face masks, are also rooted in the principles of fluid dynamics. Thus, fluid dynamics is central to all important physical aspects of the airborne transmission of respiratory infections such as COVID-19, and it therefore stands to reason that this connection to flow physics will appear in any successful model of airborne transmission. In the sections that follow, we provide additional details about the key variables involved in the CAT inequality, with special emphasis on the intervening fluid dynamical phenomena. This is followed by the application of the model to assess transmission risk associated with face mask use, physical distancing, and exercise intensity. Finally, we describe the caveats associated with this model and summarize the study.

## 4 Host Related Variables

The CAT inequality (Fig. 2) naturally segregates into three sets of variables: the first set depends primarily on the host, the second on the environment, and the third, on the susceptible. We now describe the factors that each of these variables depend on, as well as our state of knowledge regarding each variable.

$\dot{R}_h$ is the rate of expulsion of respiratory droplets from the nose and mouth of the host, and is one of the most extensively studied parameters within the arena of airborne transmission[6,7,9,25–28]. Droplets are formed from the mucus and saliva that lines our respiratory and oropharyngeal tracts, and these droplets are expelled with the air that is exhaled out of our mouth and nose. Studies have shown that individuals generate more droplets for the same expiratory activity while ill with a respiratory infection than after recovery[26,29], and this may be related to enhanced mucous production during illness. While the conventional notion is that sneezing has the highest rate of droplet generation followed by coughing, talking[7,30] and breathing[6] (in that order), the very large scatter in measured data[6,25,31] makes it difficult to validate this notion. Attention during the ongoing pandemic has focused on droplet generation during talking and breathing[7,8,25] due to the recognition that viral shedding from asymptomatic/presymptomatic individuals (who are not coughing or sneezing) may be an important differentiator in the high spreading rate of SARS-CoV-2 infections compared to earlier coronavirus outbreaks[2,32].

For activities such as breathing and talking, it could be appropriate to express $\dot{R}_h$ as the product of the volume expiration (i.e. ventilation) rate of the host, ($\dot{V}_{Eh}$) and the number density of droplets (i.e. droplets per unit volume) in the exhaled gas ($n_{dh}$). This is because for a given individual, $n_{dh}$ might not vary significantly during activities such as breathing, and $\dot{R}_h$ would therefore increase linearly with the ventilation (i.e. exhalation) rate $\dot{V}_{Eh}$. The ventilation rate for an adult can range from about 100 ml.s$^{-1}$ at rest to 2000 ml.s$^{-1}$ during intense exercise[33]. Measured values of $n_{dh}$ for breathing[34] are about 0.1 ml$^{-1}$ and this suggests that $\dot{R}_h$ for breathing could range from about 10 to 200 droplets per second depending on the ventilation rate. Values of $n_{dh}$ during normal speech in the same experiment were found to be about two to eight times higher, and other studies have found that droplet emission increases with the loudness of speech[7]. Finally, recent attention has focused on "super producers,": individuals who according to some studies, generate droplets at rates that are 10 or more times higher than others[29,35]. Thus, even for normal breathing, $\dot{R}_h$ could range from about 10 to 2000 s$^{-1}$ depending on the exhalation rate and the emission phenotype of the individual, and speech could increase the upper range by another order of magnitude. Thus, phenotype and expiratory activity of the host alone could increase the transmission risk by a factor of 1000 or more.

$f_{vh}$ is the fractional viral load of a respiratory droplet and there is currently no data on this variable for SARS-CoV-2. Indirect measures based on volume concentration of viral load in oral fluid samples collected from COVID-19 patients combined with simple volumetric estimates have been used to suggest that 37% of 50 μm size droplets and 0.37% of 10 μm size droplets would contain virions[30]. No confirmation of these estimates from direct measurement of respiratory aerosols is available so far, and there is evidence that suggests that these simple volume-fraction based estimates might significantly underestimate the viral load of the small (<5 μm) droplets[27]. Furthermore, the fractional viral load also likely depends on the location in the respiratory tract from where the droplet originates because pathogens tend to colonize specific regions of the respiratory tract and the surface area density of the mucus volume varies throughout the respiratory tract[36]. There is, however, no quantification of this effect. We note that employing even a low-end estimate of say a 0.5% fractional viral load (i.e. $f_{vh}$= 0.005), combined with 200,000 droplets/cough[26], would result in the shedding of 1000 virions in each cough, and this could be equivalent to the minimum infectious dose (see discussion in Sec. 6) for COVID-19.

$f_{ah}$ is the fraction of expelled droplets that aerosolize, i.e. get suspended in the air. It is generally found that droplets smaller than about 10 μm can remain suspended in the air whereas droplets larger than 50 μm fall to the ground rapidly[34,37–39]. Thus, the size distribution of the expelled droplets is a key determinant of $f_{ah}$. A number of studies have examined the distribution of droplet size expelled during various expiratory activities[6,7,9,25–28] and these studies show that droplet size can vary from 0.01 to 1000 microns.

The consensus is that breathing generates the smallest particles, with talking, coughing, and sneezing generating increasingly larger droplets (in that order)[40]. There is however a large scatter in this data, and this might be due to subject-specific differences[41] as well as the changes in mucosal fluid induced by the pathogen[40]. Finally, the fluid in the expelled droplets also evaporates rapidly, resulting in a reduction in size, and this rate of evaporation may depend on prevailing conditions of temperature and humidity near the host, as well as the velocity of the droplets. These dependencies can, however, be determined, for the most part, from first principles[38,42].

## 5 Environment Dependent Variables

$f_{at}$ represents the fraction of aerosolized respiratory aerosol droplets/droplet nuclei from the infected host that are transported to the immediate vicinity of the susceptible, and this is one variable where environmental factors play a dominant role. These include air currents, turbulence, temperature, and humidity. Ambient air currents in particular, determine the "time-of-flight" as well as the dilution in concentration of the bioaerosol that arrives near the susceptible.

Even though we know the dependencies of the variable $f_{at}$, it is still a difficult variable to estimate since environmental factors can be so highly variable[43]. For example, even for a host and susceptible in the same room, this variable could change significantly given the relative location of the two individuals, the operational status of the air conditioning, and the location of the individuals relative to the air conditioning diffusers and vents[15,44,45]. The estimation of this parameter becomes even more difficult in indoor spaces such as buildings where rooms share a high-volume air conditioning (HVAC) system. In high-density indoor spaces such as classrooms, aircraft cabins, gyms, buses, trains, etc., anthropogenic effects generated due to human movement and body heat generated thermal plumes[46,47] could also have a significant effect on this variable. The effects of indoor ventilation fluid dynamics on COVID-19 transmission have been recently reviewed[48].

Estimation of $f_{at}$ in outdoor environments presents a different challenge. While these outdoor environments do not have confining boundaries and localized inflow/outflow regions that dominate the flow patterns, effects due to atmospheric turbulence[49,50], local wind and weather conditions, convection effects due to thermal gradients, and other environmental factors have to be taken into account. Furthermore, even in outdoor settings, the presence of high human density (such as at sporting events, social gatherings etc.) could introduce significant anthropogenic effects on the dispersion and transport of respiratory aerosols.

As the aerosol plume from the host travels downstream, it spreads due to diffusion, entrainment, and turbulence-induced mixing. This results in a direction-dependent drop in concentration (aerosol particles per unit volume) with distance from the host. To further understand how this enters the estimation of the variable $f_{at}$, we introduce the variable $V_s$, which is the volume of air surrounding the face of the susceptible that would be inhaled by the susceptible (see Fig. 3). If the aerosol concentration near the host is $C_o$ and the mean concentration in the volume $V_s$ at a distance of $D_{hs}$ is $\bar{C}_s$, then $f_{at}$ due to this dilution in concentration can be expressed as $\bar{C}_s/C_o$. The volume $V_s$ can be estimated given the inspiratory status of the susceptible (see discussion of $f_{is}$) but in the current model, a choice for $V_s$ that eliminates dependence of this variable on the susceptible, is the maximum possible volume of air that can be inhaled by an adult per second, which is about 2 liters[33]. Thus, $f_{at}$ can be estimated under these assumptions if $\bar{C}_s/C_o$ is known or can be estimated. We point out that the 'exposure index' of Liu et al.[51] is defined in a similar way, and in their study, is estimated using computational modeling. It is also noted that $C_o$ could be expressed in term of the ratio of the particle expulsion rate to the exhalation rate of the host as $C_o = \dot{R}_h/\dot{V}_h$.

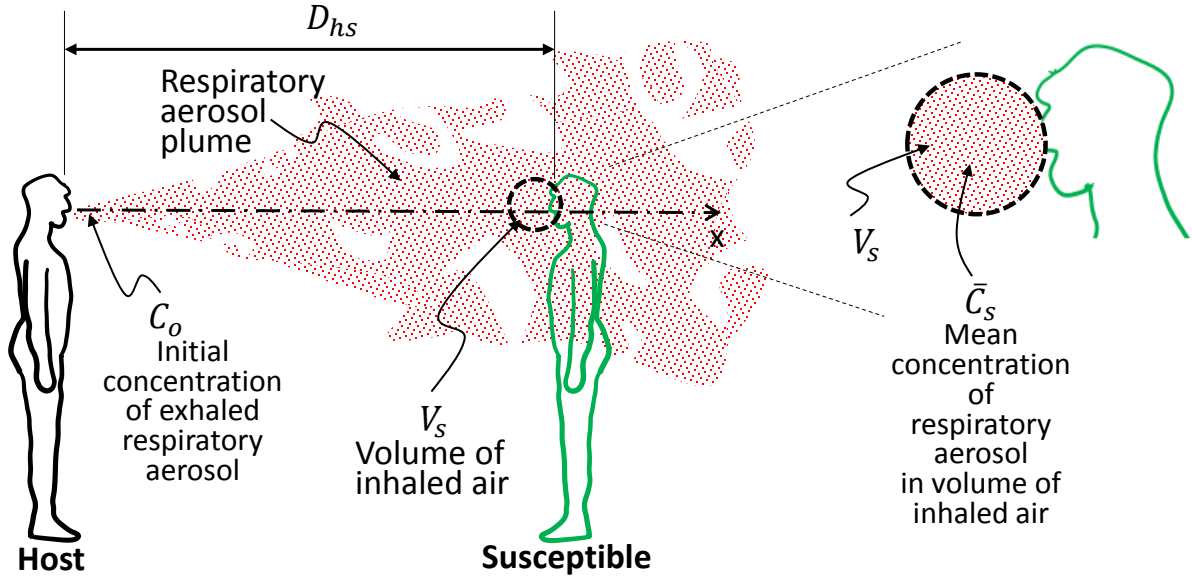

*Figure 3. Schematic depicting the inhalation volume of the suceptible that can be combined with the local concetration of the respiratory aerosol to estimate $f_{at}$.*

A number of studies have measured the spreading rate of the exhalation jets formed from various expiratory activities [52–54] and this spreading rate with downstream distance can vary significantly for breathing, talking and coughing. External flow currents and thermal convection[33] effects, will however deform the shape of the expiratory plume and may enhance non-uniformity in the concentration of the respiratory aerosols within the jet. Quantifying the effect of all these factors is the key challenge in estimating $f_{at}$. In Sec. 8, we will employ canonical computational models of scalar dispersion to provide estimates of the protection factor associated with physical distancing in several scenarios.

$f_{vv}$ represents the fraction of respiratory aerosol particles from the infected host that arrive in the immediate vicinity of the susceptible *with viable virions*. Ambient air currents determine the "time-of-flight" of the aerosol particles, and this combines with temperature, humidity and UV exposure to determine the viability of the virions carried in the aerosols. A study has shown that the SARS-CoV-2 virion can stay viable in aerosol form for 3 or more hours[13], but high temperature[55] and sunlight/UV exposure[56,57] are both detrimental to virion viability. Humidity, on the other hand, has a more complex effect on the viability of airborne viruses. For instance, three regimes of Influenza A virus viability in droplets, defined by relative humidity (RH), have been postulated: high viability at physiological (~100% RH) and dry (< 50% RH) conditions, and lower viability when intermediate humidity (50% - 100% RH) exists in combination with high concentrations of naturally occurring biochemical solutes in the droplet[58]. This complex dependency on humidity is likely one factor that has made it difficult to correlate transmission risk with regional and seasonal variations in environmental conditions[59]. In general $f_{vv}$ can be modeled as $e^{-T/\tau}$ where $T$ is the time-of-flight and $\tau$ is the half-life of virions in aerosolized form, which depends on the virus, as well as the temperature, humidity and UV exposure. A number of recent studies have measured $\tau$ for SARS-CoV-2 in a variety of settings[13,56].

## 6  Susceptible Related Variables

$f_{is}$ is the fraction of bioaerosols from the host in the vicinity of the susceptible that would be inhaled and deposited in the respiratory tract of a susceptible not wearing a face covering. This variable primarily depends on the inspiratory status of the susceptible. At rest, an adult human inhales about 100 ml of air per second[60] but this value can go up 20 fold during intense exercise[33]. Within the context of the current model, $f_{is}$ could be estimated as the ratio of the susceptible's inspiratory rate to the maximum possible inspiratory rate for a human (we denote this maximum ventilation rate as $\dot{V}_{max}$.), which can be assumed to be 2000 ml/second[33]. We note that the volume here is the same as $V_s$ in the previous section. With this prescription, $f_{is}$ for the average healthy adult male could vary from 0.05 during rest to 1.0 during intense exercise. Beyond the exercise state of the individual, tidal volume (volume inhaled per breath) and ventilation rate also depend on age[61], gender, body weight[62,63] and the respiratory health of the person, and these factors can be easily accounted for in $f_{is}$. For instance, measured values of ventilation rates for women are about 20% lower than for men[62] and this would translate to a 20% reduction in $f_{is}$ for women. Similarly, short adults can have resting inspiratory rates that are about 20% lower than tall adults[62] and this would result in a proportionate reduction in $f_{is}$. Inspiration rates for preteens can be three-fold lower than adults[61] and this would also reduce $f_{is}$ proportionately. Thus, differential inspiration rates could play a role in the age and body-weight associated COVID-19 prevalence disparities noted in recent studies[64,65]. The effect of physical activity-associated changes in ventilation rates on transmission risk is examined in Sec. 8.

$N_{ID}$ is the infectious dose for airborne transmission. In the arena of infectious diseases, the infectious dose is often expressed as $HID_{50}$[22], which is the minimum infectious dose required to initiate infection in 50% of inoculated humans. This number is usually obtained via controlled studies where human volunteers are exposed to different viral loads. However, such studies are not available for potentially lethal viruses such as SARS-CoV-2. Studies on the infectious dose for Influenza A indicate a $HID_{50}$ of $O(1000)$ virus particles[22,66]. Studies of MERS-CoV in mice found a similar infectious dose[67], so the $HID_{50}$ for humans accounting for the larger body weight, could be two or more orders-of-magnitudes higher. It is important to note that for Influenza A, infectivity via aerosols has been found to be $O(10^5)$ higher than via a nasopharyngeal (i.e. nasal swab) route,[68] highlighting the exceptional effectiveness of the airborne route for transmission of respiratory infections. Infectivity of airborne viruses also depends on the carrier droplet size. Small (~2 µm) droplets deposit deeper in the lungs and have been shown to be two or more orders of magnitude more infective than larger (>10 µm) droplets[69]. Finally, the infectious dose might also depend on the age and health status (including the level of immunity to infection) of the susceptible. Determination of $N_{ID}$ for SARS-CoV-2 remains one of the most important tasks for scientists working in this arena.

The remaining variable $T_s$ is the duration of exposure of the susceptible to the aerosols from the host. Based on the CAT inequality, if all other conditions remain stationary, the total number of viable virions transmitted is directly proportional to the exposure duration. Transmission is successful when this viral dose equals or exceeds the infectious dose (i.e. when $T_s \geq T_{sC}$).

## 7  Face Coverings

Face coverings appear in the two factors $f_{mh}$ and $f_{ms}$ as fractions of aerosols/droplets that pass through the face coverings of the host and susceptible, respectively, and there is much data available to estimate these variables. These face covering-related variables depend on two factors – the material of the face covering and the fit of the face covering on the face of the individual. A perfectly fit N95 face mask for instance, stops 95% or more of the particles that go through it and $f_m$ would therefore be equal to 0.05 for such a mask. Thus, the wearing of a well-fit N95 mask by either the host or the susceptible could reduce the transmission risk by a factor of 20. Furthermore, if both individuals are wearing such masks, the transmission risk, according to the CTA inequality, could reduce by a factor of 400.

Surgical masks have been measured to block 30% to 60% of respiratory aerosols[70,71] and in another study, a surgical mask reduced aerosol shedding of Influenza A virions from infected hosts by a factor of 3.4.[28] This suggests that even surgical masks worn by both the host and the susceptible could reduce transmission risk by factors ranging from about 2 to 10. Another recent study of viral shedding with and without surgical face masks from patients with Influenza, coronavirus (SARS) and rhinovirus provides clear evidence of the ability of such face covering to reduce transmissibility of the virus[72]. Finally, we point out that even home-made cloth masks provide some protection against airborne infections[71], and in Sec 8, we will examine this in more detail for a range of fabrics.

The fitment of the mask is important for overall protection since a loose-fitting mask with perimeter leaks allows unfiltered aerosols to bypass the mask[73]. Leaks are a particular problem for outward protection (i,e, reducing emission of respiratory aerosols by the infected host) since the process of expiration pushes the mask outwards and enhances perimeter leaks[11,74,75].

Finally, in addition to filtering aerosol particles, face covering also reduce the velocity of exhalation jet[74,76] This could increase the expansion angle of respiratory jet and reduce the initial penetration distance of the respiratory droplets,[77] thereby altering $f_{at}$. For instance, a recent study found that the fabric in some face coverings might facilitate the breakup of large droplets into smaller ones[78], and in doing so, increase $f_{at}$, thereby increasing transmission risk.

## 8  Model Predictions

The model is now applied to address three distinct questions: what protection is afforded by different face coverings; how does risk decrease with increased physical distance between the host and susceptible, and finally, to what degree does the level of physical activity, as manifested in the ventilation rates of the host and/or the susceptible, affect transmission risk.

<u>Protection Afforded by Face Coverings</u>: We start with the effect of facemasks and employ data from Zangmeister et al. [71] on the filtration efficiency (*FE*) of common fabrics used in respiratory face masks. These authors examined more than 30 different fabrics and quantified the filtration efficiency for droplet sizes ranging from particle mobility diameters between 50 and 825 nm. Given that aerosol transmission may involve droplet sizes ranging up to 5 µm, we have estimated the lower and upper bounds of the average filtration efficiency ($\overline{FE}$) for particle sizes ranging from 50 nm to 5 µm (see procedure in the Appendix). The fraction of aerosols/droplets that pass through the face coverings is then given by $f_m = 1 - \overline{FE}/100$. Given these values and the assumption that the filtration efficiency is the same for inward as well as outward protection (i.e. $f_{mh} = f_{ms} = f_m$) we can now estimate the unilateral protection factor (*PF*) if either the host or the susceptible wears this mask, as *PF* = $f_m^{-1}$. The corresponding bilateral protection factor, i.e. when both individuals wear masks, is then given by *PF* = $f_m^{-2}$. These PFs normalized by the corresponding situation where neither individual is wearing a mask, are plotted in Fig. 4 for selected cases from Zangmeister et al[71].

Given the large uncertainties in estimates of the protection factor derived here, only general conclusions regarding the face masks are drawn here. First, a number of simple fabrics (Cotton 4, Cotton 14, and Synthetic Blend 2), provide protection factors that are similar or better than the Surgical Mask sample. Second, even the lower bound of unilateral protection for many of these samples exceeds 2.0, which represents a significant reduction in transmission risk. Third, for bilateral protection, Cotton 4 and Syn Blend 2 have minimum protection factors that exceed 5. Finally, the true protection factors for these fabrics are likely significantly higher than the lower bounds established here. Indeed, if we average the upper and lower bounds of $\overline{FE}$ for the four most effective fabrics/samples (Cotton 4, Cotton 14, Synthetic Blend 2, and Surgical Mask), we obtain an aggregate filtration efficiency of 63%, which would correspond to a unilateral(bilateral) protection factor of 2.7(7.3). Thus, simple facemasks made from any of these fabrics/materials could significantly lower overall transmission rates. This effectiveness of facemasks is being corroborated by recent epidemiological[16] and animal[79] studies of COVID-19 transmission. We point out that the above analysis ignores perimeter leaks, which can significantly deteriorate the effectiveness of face masks[11]. The analysis also does not account for unsteady and velocity dependent effects of expiratory events on the filtration efficiency[74].

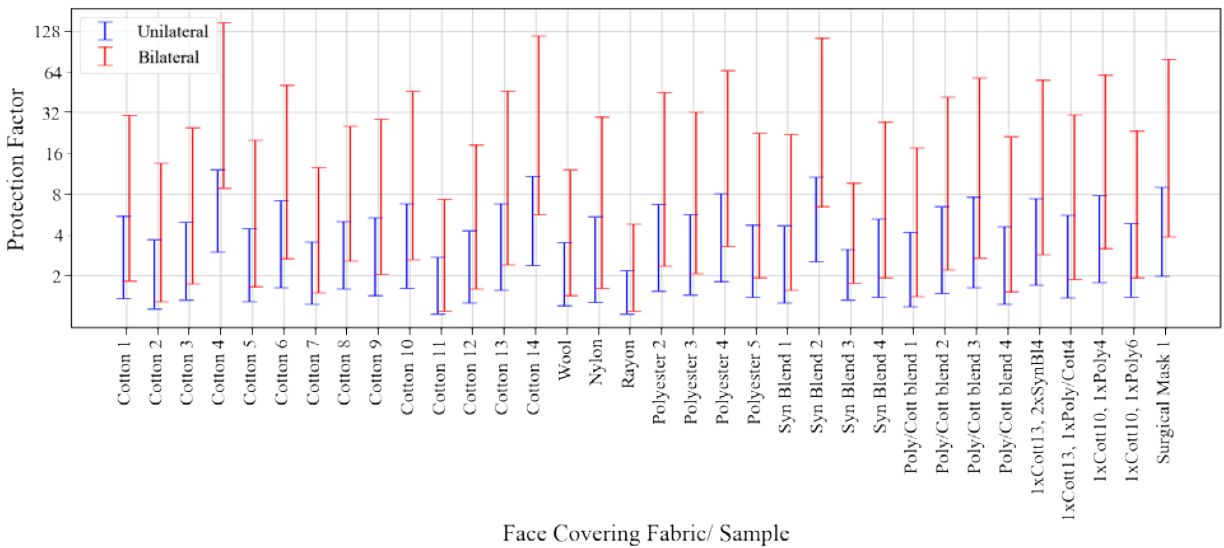

*Figure 4. Estimation of protection from aerosol transmission afforded by the donning of face coverings based on published filtration efficiency data of Zangmeister et al.[71] for 34 different fabrics/samples. The protection factor (PF) is a quantity that is normalized by the risk of transmission associated with the situation when neither the host nor the susceptible wears a mask.*

Protection due to Physical Distancing: The model is used next to examine the protection from transmission afforded by physical distance between the host and the susceptible in an outdoor environment. As mentioned earlier, the distance between the host and the susceptible is a dominant factor in the variable $f_{at}$ associated with the transport of virion-bearing aerosols. However, estimates for the rate at which transmission risk diminishes with distance between the host and the susceptible, are not readily available. Here, we employ simple models of the aerosol dispersion to estimate the protection factor associated with physical distancing.

We start by assuming that the mean concentration in the inhaled volume of the susceptible is equal to $C_{max}(D_{hs})$, where $C_{max}$ is the maximum concentration at any given distance from the host (see Fig. 5). Dispersion-induced dilution at a distance of $D_{hs}$ would then result in $f_{at} \sim C_{max}(D_{hs})/C_o$, and a corresponding protection factor due to physical distancing of $f_{at}^{-1}$. Several scenarios for outdoor transmission are considered based on various combinations of expiration velocity ($V_j$), ambient wind speed ($U_\infty$) and buoyancy induced effects. We note that these models are consistent with the approach inherent to the CAT-Inequality which assumes a sequential segregation of the various effects involved in airborne transmission. In particular, important near-field effects such as droplet breakup and evaporation are assumed to be accounted for in the variable $f_{ah}$ and are therefore not included in the droplet dispersion models discussed in the current section.

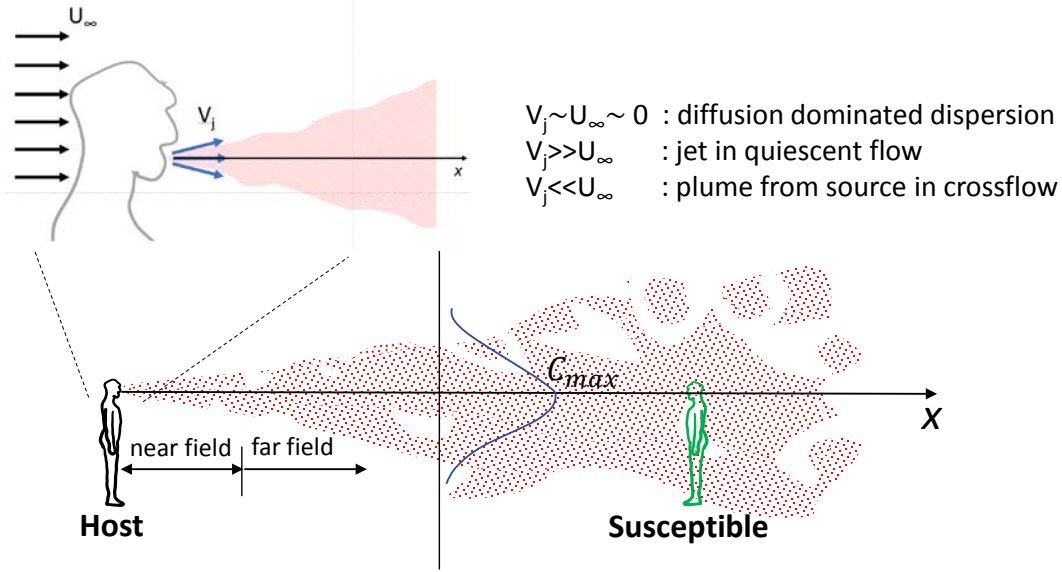

Figure 5. *Schematic showing various scenarios for which the effect of physical distancing on transmission risk is assesed. The aerosol plume from the host consists of a near field, which can be highly variable, and a far field, where the plume exhibits more self-similar or universal characteristics within various classes of flow. The analysis in this section focuses on the far-field domain.*

1) $V_j$ and $U_\infty$ are both of very low magnitudes; this could corespond to a sedentary individual breathing at a low exhalation rate in still wind conditions. This situation can be modeled as normal diffusion dominated dispersion, and the analytical solution for steady-state diffusion from a point source[80] in an unbounded domain indicates that $C_{max}(x) \sim x^{-1}$ where x is the distance from the point source. Molecular diffusion will not be relevant since it will take a long time to establish itself (on the order of $t \sim x^2/\gamma$, where $\gamma$ is the molecular mass diffusion coefficient, typically many hours in air). However, even relatively weak background turbulence with eddies smaller than x will generate turbulent diffusion coefficients[81] $\gamma_T >> \gamma$ and also establish a $C_{max}(x) \sim x^{-1}$ spatial decay.

2) $V_j$ significantly exceeds $U_\infty$; this could correspond, for instance, to a person talking or singing (where expiratory flow speeds range up to 5 m/s [54]) in still wind conditions. This situation could be approximated as a turbulent jet in quiescent flow, and studies[82] of such flows in canonical configurations indicate that the peak concentration decays beyond the near field as $C_{max}(x) \sim x^{-1}$ in the direction of the jet. This situation has been analyzed recently for speech-driven aerosol transport including time-dependence[83].

3) $U_\infty$ significantly exceeds $V_j$; this could correspond, for instance, to a person breathing normally with an expiratory velocity of ~1 m/s[84] on a windy day with wind velocities upwards of 10 miles per hour. Neglecting bouyancy effects, this situation can be modeled as a horizontal plume from a point source in a crossflow, studies[85,86] suggest that $C_{max}(x) \sim x^{-3/2}$ beyond the near-field region.

4) The previous situation of a horizontal plume does not account for bouyancy effects and the data employed does not account for the time-dependent pulsatile nature of breathing. These effects can be included in the current model if data from appropriate computations or experiments, are available. Here we employ data generated from a wall-modeled large-eddy simulation[87] of a plume from a point source located 1.5 m above the ground in a turbulent atmospheric boundary layer, with a mean wind velocity corresponding to 2m/s[83]. The model is designed to mimic normal breathing with the scalar (representing the respiratory aerosol) being released as puffs at regular intervals of 3 seconds. The exhaled breath is assumed to be at a temperature of 37º C and two ambient temperature conditions are considered: 0º C and 42º C. In the model, buoyancy effects are included using the Boussinesq approximation. The incoming wind flow itself is assumed to be unstratified, neutrally buoyant. The reader is referred to Appendix B for details about the methodology of the simulations.

Fig. 6a shows a plot of the instantaneous scalar concentration for the second case, and Fig 6b and c, the time-averaged plume concentrations for both cases. As expected, the plume rises for the first case (the 'light' plume) but descends towards the ground for the second case ('heavy' plume) due to buoyancy effects for the very hot surrounding air case, into which the cooler and denser air is exhaled. Fig. 6d shows the maximum concentration of the respiratory plumes as a function of distance for the two cases, and we find that beyond about a distance of about 3 meters, the plume concentration decays consistently as $C_{max}(x) \sim (x^{-1.2})$ and $C_{max}(x) \sim x^{-0.9}$ for the light and heavy plumes, respectively. Thus, the presence of the ground as well as buoyancy has a noticeable effect on the concentration decay rate.

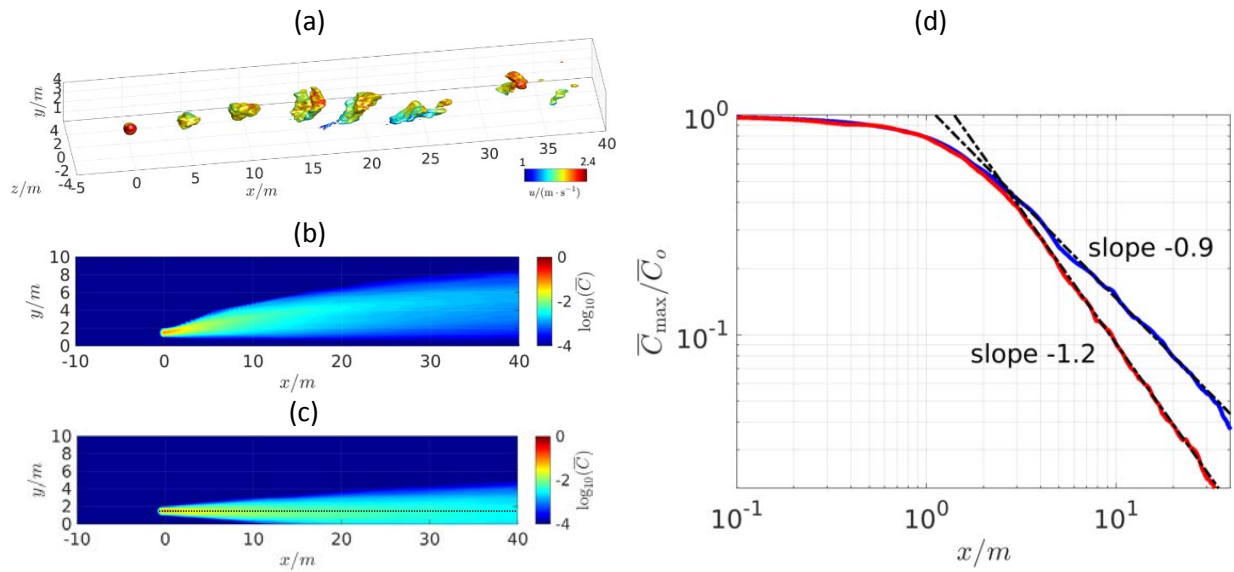

*Figure 6. Results from wall-modeled LES of a breath generated aerosol plume (at x=0) in a turbulent boundary layer.(a) isosurfaces of instantaneous concentration of scalar $C/C_o$=0.01, colored by the local streamwise velocity showing the breath aerosol puffs being transported in the turbulent flow (b) contours of the mean concentration for plume that is warmer than the ambient flow (c) contours of the mean concentration for a plume that is colder than the ambient flow. (d) Mean concentration with streamwise distance (in meters) at a height of 1.5m from the ground along with best fit power laws beyond the near-field region.*

Fig. 7 shows the protection factors due to physical distancing for all the cases discussed above and we note that since the y-intercepts of all the curves have been individually normalized to unity at a unit distance, direct numerical comparison between the two conditions is not appropriate. The plot does however indicate that in the absence of a crossflow, the protection factor increases linearly with distance, whereas when the crossflow velocity is significantly larger than the exhaled jet velocity, the protection factor increases at a faster rate of $D_{hs}^{1.5}$. Furthermore, buoyancy has a noticeable effect on the decay of aerosol concentration, and the downward movement of heavy plume combined with the confinement due to the ground, could potentially diminish the protection afforded by physical distancing. Given that wind conditions can be highly variable, a conservative estimate from the above analysis is that physical distancing affords approximately a linear increase in protection from transmission. The analysis also demonstrates the use of data from computational fluid dynamics models to parameterize the model for specific scenarios.

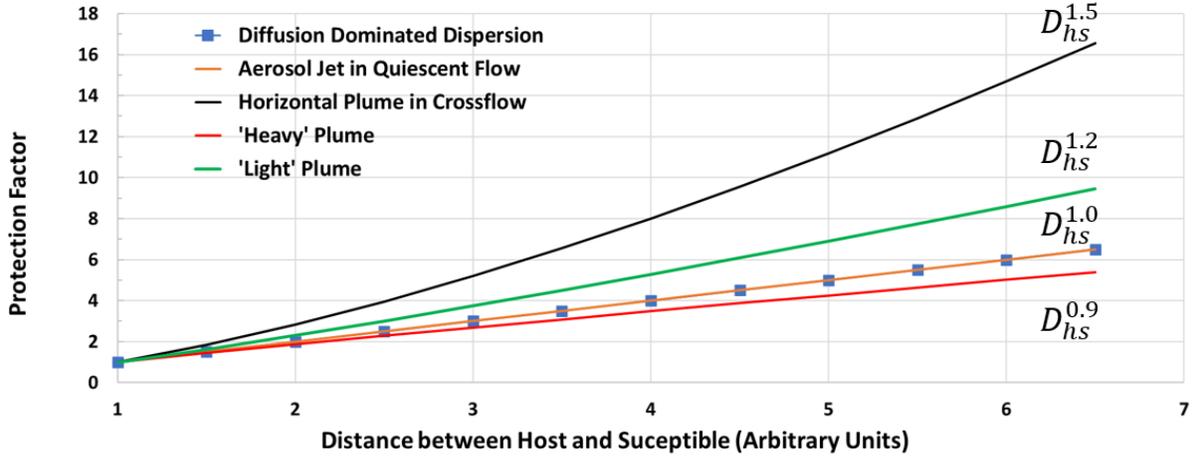

*Figure 7. Estimate of protection factor as a function of physical distance (in arbitrary units) between the host and susceptible for the five scenarios examined here. The protection factor is normalized for each case by the condition where both the host and susceptible are at a unit distance. The protection factor is inversely proportional to the decay of the maximum concentration with distance between the host and the susceptible $D_{hs}$.*

Level of Physical Activity and Transmission Risk: The final application of the model is to examine the potential effect of physical/exercise intensity on the risk of transmission. This would be relevant to settings such as gyms, sports/exercise facilities, and even gatherings/events, schools, and workplace situations where levels of physical activity might exceed levels that are considered sedentary. Exercise intensity enters the CAT inequality through the ventilation rates of the host and the susceptible. As pointed out earlier, in expiratory activities such as breathing and talking, the particle expulsion rate may be estimated as $\dot{R}_h = n_{dh}\dot{V}_{Eh}$ where $n_{dh}$ is the number of droplets emitted per volume of exhaled gas (and may be assumed to be constant for a given host) and $\dot{V}_{Eh}$ is the ventilation rate of the host. The variable $f_{is}$ is equal to the rate of the susceptible's ventilation rate divided by the maximum possible ventilation rate i.e. $f_{is} = \dot{V}_{Es}/\dot{V}_{max}$. Employing established definitions[88] that relate exercise intensity to oxygen consumption rates, and further assuming proportionality between oxygen consumption rates and corresponding ventilation rates, and that the maximum ventilation rate $\dot{V}_{max}$ for adults is 2 liters/sec[89], we can estimate the increased transmission risk with exercise intensity over the sedentary condition as $(\dot{R}_h f_{is})/(\dot{R}_{h,min} f_{is,min}) = \dot{V}_{Eh} \times \dot{V}_{Es} / \dot{V}_{min}^2$, where $()_{min}$ corresponds to an adult in a sedentary condition. In the current estimation procedure, $\dot{V}_{min}$ is set at 100 ml/s, which is 5% of the maximum ventilation rate, and the ventilation rates for the intermediate levels are based on measured values for adults[88].

Fig. 8 shows this increased transmission risk for the five exercise intensity levels and it can be seen that even for a susceptible in a sedentary state, the transmission risk goes up by a factor of 8 if the host is at a moderate intensity of exercise. In a settings such as shopping malls, outdoor markets, warehouses or high-schools, where activity levels of hosts and susceptibles could be in the light to moderate range, increase in transmission risk due just to increased ventilation rates, would, according to the current model, be up to 64 times higher. In a facility such as a gym, or for instance a basketball practice, where exercise intensity levels could be in the "vigorous" range, transmission risk could be nearly 200 times higher due to increased exhalation and inhalation rates of the individuals involved. A survey of existing literature indicates that this increase in risk with physical activity level associated with various common scenarios is under-appreciated.

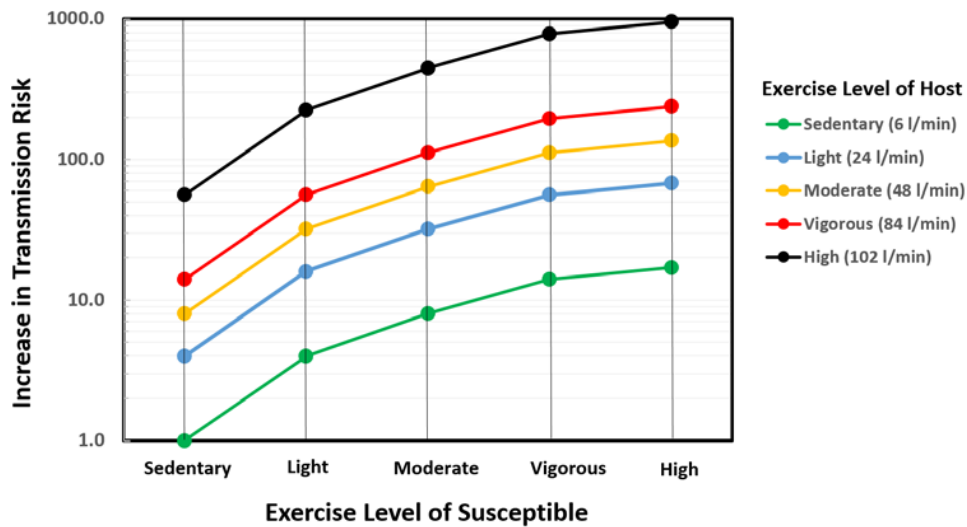

*Figure 8. Estimate of transmission risk increase due to physical activity/exercise induced increased ventilation rates for host and susceptible. The assumed ventilation rate for each of the 5 levels is included in the legend. The increase in transmission risk is normalized by the condition where both the host and susceptible are sedentary (sleeping or sitting).*

## 9  Comparison with other Infection Risk Models

Existing models for estimating infection risk via airborne transmission can be classified into Wells-Riley type and dose-response models[90]. Within this context, the current model should be considered a dose-response model since it explicitly makes use of the infectious dose ($N_{ID}$) predict infection risk. The current model, as proposed, can further be characterized as a deterministic (as opposed to stochastic) dose-response model since it assumes that a dose larger than the threshold dose of $N_{ID}$ results in an infection[90].

The Wells-Riley type models are based on the notion of an infectious "quantum" which is defined as the quantity of expelled aerosol required to cause an infection in a susceptible[24,21]. Within the context of the CAT inequality, the quanta emission rate can be expressed as $(\dot{R}_h f_{vh}/N_{ID})$ and the CAT inequality could easily be reformulated to express risk in terms of this quantity. However, the quanta emission rate combines a host-dependent variable (the rate of viral shedding) with a susceptible dependent variable (the infectious dose) and makes it difficult to delineate the effects of the distinct states (health, inspiratory state, mask use, etc.) of the two individuals involved. Furthermore, the vast majority of such models assume a "well-mixed" state for the aerosols in the environments and this does not allow for "local" effects[90] that have direct bearing on practices such as social distancing.

Dose-response models of varying degree of complexity have been developed[90]. Many of these models allow for spatial and temporal inhomogeneities in the aerosol concentrations and are well-suited for detailed modeling of infection risk in a variety of scenarios. Some of the recent models that have been developed can incorporate data on ambient flow conditions from computational models or experimental measurements[91]. However, such models are expressed in mathematically complex forms, which diminishes comprehensibility outside disciplinary expertise, and makes it particularly difficult to communicate the underlying ideas to non-scientists. As shown in the previous sections, the simple mathematical form of the current model and its phenomenology-based compartmentalization into host, environment and susceptible dependent variables, not only allows for easier comprehension by a wide range of audiences, but also provides quick estimates of factors including, but not limited to, the type of mask worn, physical distancing, and the inspiratory status of the host and susceptible.

Finally, we point out the left-hand side of the CAT equality represent the total aerosol viral dose inhaled by the susceptible, and if we denote this variable by $N_{Ds} = \dot{R}_{tot} T_s$ , then $\left(\frac{N_{Ds}}{N_{ID}} - 1\right)$ represents the normalized viral 'overdose' delivered to the susceptible. Thus, in addition to evaluating the risk of transmission, the CAT inequality can also be used to assess the degree of exposure of the susceptible to aerosolized virus, which is known to be correlated with the severity of the infection[92].

## 10 Caveats

The notion that "a model is a lie that helps us see the truth[2]," certainly applies to the current model as well. The CAT inequality is an attempt to express the highly complex, multifactorial process of airborne transmission of a respiratory infection such as COVID-19 in a simple way, and the following caveats and limitations of this model are worth pointing out:

1. The choice of the variables in the CAT inequality is not unique and other combinations of the variables are possible. In particular, the variables shown in the CAT inequality could be decomposed further; for instance, $\dot{R}_h$ can be expressed as the rate of droplet generation in the respiratory tract and the fraction of generated droplets that are expelled from the mouth. Such a variable separation might be appropriate, for instance, to isolate the effect of therapies that attempt to diminish the droplet generation rate via alteration of the mucous properties[35].
2. The inequality assumes that the rate of arrival of virion-bearing aerosols in the vicinity of the susceptible is constant in time. The recent analysis[83] of speech driven aerosol transport takes into account the start time and travel duration in scenario (2) treated in section 8. The CAT inquality could be modified to include a time-dependent emission and arrival rate[21] but this would increase the complexity of the mathematical expression. Assuming steady state condition results in predictions that are more conservative in most cases. Finally, it is also assumed here that the rate of virion arrival (i.e. proportional to the advective or diffusive flux of *C*) in the vicinity of the susceptible is sufficiently high so that *C{max}* there is not markedly affected by the inhalation process itself. If the flux is not high enough, the inhalation may deplete *C{max}* near the susceptible over time. The analysis for flux-limited situations again introduces additional complexities.

---

[2] A quote attributed to Howard Skipper, an American doctor.

3. The CAT inequality could be missing important, but as yet unknown effects. For instance, the use of $N_{ID}$ in the model assumes that it is the accumulated dose of virus that determines transmission. While this assumption is quite standard in the arena of infectious diseases[21–23] it is plausible that the rate at which this infectious dose is delivered to the respiratory tract of the susceptible is also important in initiating an infection. For instance, 1000 virions inhaled over a short duration (say minutes) might overwhelm the immune system whereas the same viral dose delivered over a much longer duration (say hours) might allow the immune system to mount an effective response and avoid infection.
4. The variables in the CAT inequality are more accurately represented as variables with probability density functions (PDFs) given the stochastic nature of the processes involved[93]. For instance, respiratory droplets of different sizes are expelled at different rates[9,28,31,36] during an expiratory event and the rate of droplet emission $\dot{R}_h$ could therefore be expressed as a droplet size-dependent PDF. Similarly, the viral loading of respiratory droplets ($f_{vh}$) as well as the infectious dose $N_{ID}$ are expected to be functions of droplet size, and could therefore be represented by droplet size dependent PDFs. The often turbulent environment determining $f_{at}$ is expected to cause significant fluctuations in travel time, turbulent diffusion rates and individual eddying events can influence the local concentrations. Hence, the factor $f_{at}$ itself has a mean value as well as a distribution around that mean value that depends on detailed flow conditions.
5. If the variables in the CAT inequality are interpreted as random variables with PDFs (as in 4. above), it implicitly assumes the variables and factors are statistically independent. However, variables in the CAT inequality are not necessarily mutually independent given the fact that many of them have common dependencies. The joint dependency of many variables on particle size has already been described above. Other examples include the face covering on the host which modifies $f_{mh}$. A resulting alteration of the expiratory jet due to the mask could also affect the aerosolization variable $f_{ah}$ of the expelled droplets as well as entrainment into the ambient air current, which could affect $f_{at}$.
6. The model assumes a single host but the CAT inequality can easily account for multiple hosts by summing the left hand side for multiple infected hosts.
7. Validation of the proposed model is not attempted here. Validation requires accurate estimates of inputs as well as outputs to the models, and these are difficult to obtain, especially for potentially lethal infections such as COVID-19. Indeed, most models of airborne infection to date, including classic[24] as well as more recent models[21,94–96], remain un-validated. Despite this lacuna of validation, the value of all such models, including the current one, is that they enable an examination of how transmission risk scales with key variables, and an assessment of the effects of mitigation strategies on overall transmission rates.

## 11 Summary

The CAT Inequality is a mathematical model for estimating the risk of airborne transmission of infectious diseases such as COVID-19, that is expressed in a simple and intuitive way so as to convey the factors involved in transmission to a wide range of stakeholders ranging from scientists from various disciplines, to policy makers, public media and even the general public. As shown through specific examples, the model provides a framework for interpreting and quantifying the relative changes in risk from behaviors such as, but not limited to, wearing masks, physical distancing, and the intensity of physical activity/exercise on infection risk, in terms that are easy to convey to a range of audiences. The approximately inverse relationship of transmission risk and spatial distance from physical distancing is one example of the important insights that can be generated by the model.

In closing, we point out that while the transmission model presented here is inspired by the Drake Equation, the current model is not speculative but of a more deterministic nature. This is because we understand much more about the factors involved in this transmission model than we do about the factors in the Drake equation. Indeed, as discussed in the paper, estimates for many of the variables in the CAT inequality can be obtained from existing data or from basic principles of fluid dynamics, physiology, and virology. Even for the variables for which we currently do not have good estimates, we understand the underlying dependencies as well as the procedures/methods required to estimate these variables, and it is expected that ongoing studies will close these gaps in our understanding, and provide better quantification of all the variables involved in this model.

## 12  Acknowledgements

The authors would like to acknowledge Drs. Sanjay E. Sarma (MIT), Howard Stone (Princeton University), Yuguo Li (The University of Hong Kong)and Jae Ho Lee (The Johns Hopkins University), and Raina Mittal (University of Pennsylvania) for providing feedback on a draft of this article, and Shantanu Bailoor and Ashvin Vinodh (The Johns Hopkins University) for help in preparing some figures for this paper. Data on filtration efficiencies of fabrics provided by Dr. Christopher Zangmeister (CZ) and James Radney from NIST, as well as insights from CZ on filtration physics, are gratefully acknowledged. Partial support for this work came from NSF grant CBET-1738918 and CPU time was provided by XSEDE grant TG-ATM130032.

## 13  Appendix A – Estimation of Upper and Lower Bounds for Average Filtration Efficiency of Face Mask Fabrics

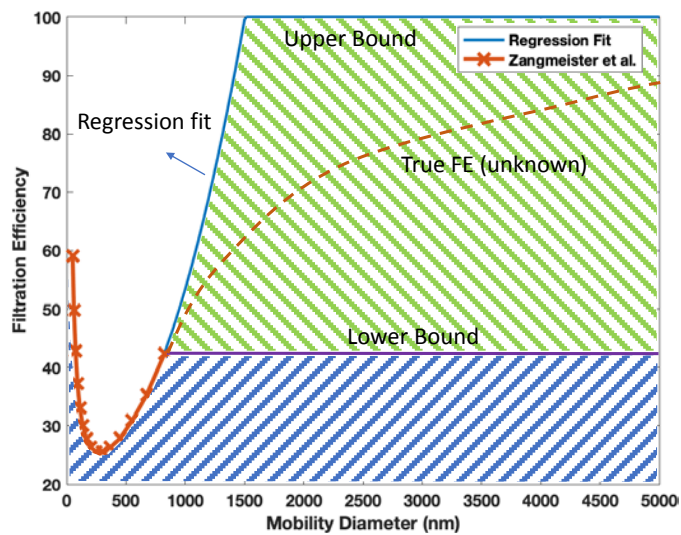

*Figure 9. Depiction of methodology for estimating lower and upper bound of the average FE over the 50-5000 nm range from the data of Zangmeister et al.[71]. This particular example is for the 1 x Cotton13, 2x Syn Blend 4 sample. The lower bound is obtained by assuming that the FE for PMDs greater than 825 nm is equal to the FE measured at 825 nm. The upper bound is obtained by employing a regression fit to the right side of the measured curve (beyond the PMD for minimum FE) and extending it to the FE=100% line. Beyond this, the FE for the upper bound is assumed to equal 100%.*

The procedure for estimating the lower and upper bounds of the filtration efficiency (FE) for various fabrics in Fig. 4 is described here. Zangmeister et al.[71], measured the FE for various fabrics as a function of particle mobility diameters (PMD) with diameters ranging from 50 to 825 nm. However, it is well established that aerosol transmission occurs with aerosol particles with diameters up to at least 5 μm. In order to apply the data from Zangmeister et al.[71] in a way that is relevant for airborne transmission, the FE versus PMD data has to be extended up to 5 μm. However, FE dependence on PMD is affected by the filter diameter, the solidity (1-porosity), the density of the particles, and face velocity and simple extrapolations of the FE versus PMD curves are not possible. Given this, and the fact that FE for any given material approaches 100% at high enough PMD, we derive a lower and upper bound of the average FE over the 50-5000 nm range as follows (see Fig. 9): the lower bound is obtained by assuming that the FE for PMDs greater than 825 nm is equal to the measured FE at 825 nm. The upper bound is obtained by employing a regression fit to the right side of the measured curve (beyond the PMD for minimum FE) and extending it to the FE=100%. Beyond this, the FE for the upper bound is assumed to be equal to 100%. A quadratic regression fit results in a $R^2$ value exceeding 0.90 for all cases. For each bound, the average FE (denoted by $\overline{FE}$) over the PMD range 50-5000 nm can now be estimated using numerical integration. Fig. 9 shows these bounds for one case cases and Table 1 shows the $\overline{FE}$ computed in this way for the various fabrics considered in the current study. Finally, the table 1 provides a brief description of the various fabrics/samples from Zangmeister et al.[71] that are included in the current study, along with the estimated lower and upper bounds of the average filtering efficiencies.

## 14 Appendix B – Wall-Modeled Large Eddy Simulation of Breath Aerosols in Turbulent Flows

Transport of periodically released breath puffs in a turbulent open channel is simulated by wall-modeled large eddy simulation (LES). The computational domain is 60 m, 10 m and 20 m in the streamwise (x), wall-normal (y) and spanwise (z) directions respectively with 672 x 117 x 384 grid points. The point source corresponding to the host is placed 1.5 m above from the ground, 10 m downstream of the inflow (defined as x=0) and at the middle span. The Reynolds number based on the bulk velocity (2 m/s), the emission height (1.5 m) and the viscosity of air, is 2.5 million. The grid has a uniform spacing of 5 cm for y⩽3m and -5m⩽x⩽10m, as well as in the z direction. It is stretched in the (x-y) plane for y>3m, x<-5m and x>10m with a maximum stretching ratio of 4%.

The number of infectious respiratory aerosol and the temperature fluctuation around the ambient temperature are modeled as two transported scalars fields. The droplet evolution is represented using the Eulerian approach where instead of solving the Lagrangian evolution of individual drops, their concentration is treated as a passive scalar. This approach can be far less expensive numerically and is justified at the low volume fractions[34] and negligible terminal velocities associated with the small aerosolized droplets whose time-evolving concentration field, the simulations aim to represent. See Refs. 97–99 for prior developments and applications of the fast Eulerian two-fluid approach. The release of the virus-laden puffs is implemented as a localized volume source term in the scalar transport equation. Buoyancy due to temperature variations is modeled in the momentum equation by the Boussinesq approximation as a function of the temperature. Subgrid-scale contributions for both the momentum are modeled by the Lagrangian-averaged dynamic eddy-viscosity mode[100,101]. For the scalars, constant subgrid-scale Schmidt and Prandtl numbers Sc=Pr=0.4 are used. Both the velocity and scalars are periodic in the spanwise direction. In the streamwise direction, the flow is periodic, while the scalars have a uniform Dirichlet boundary condition at the inflow and a convective condition at the outflow. A sponge layer is applied near the outflow to force the scalars to be zero and let the flow recovers to a canonical TBL. For both the velocity and scalar, an equilibrium rough wall model with $z_o$ = 0.5 mm is applied at the bottom wall and zero vertical gradient condition at the top boundary. The value of the scalar at the wall is the same as it in the incoming air at the inflow. The equations of motion and scalar transport are solved using second-order accurate (in time and space) derivatives on a staggered mesh. The code is parallelized using MPI and has been widely validated in similar types of turbulent flows[87,102]. Statistics are collected over 30 minutes of physical flow time after the statistical steady state is reached. Increasing the number of grid points by 26 percent in each direction (i.e., doubling the total number of grid points) has a negligible effect on the flow statistics and the results presented in the main text.

## 15 Disclosures

There are no disclosures to report.

## 16 Data Availability Statement

The data that support the findings of this study are available from the corresponding author upon reasonable request.

## 17 Bibliography

(1) Morawska, L.; Cao, J. Airborne Transmission of SARS-CoV-2: The World Should Face the Reality. *Environ Int* **2020**, *139*, 105730. https://doi.org/10.1016/j.envint.2020.105730.

(2) Morawska, L.; Milton, D. K. It Is Time to Address Airborne Transmission of COVID-19. *Clin Infect Dis*. https://doi.org/10.1093/cid/ciaa939.

(3) Lewis, D. Mounting Evidence Suggests Coronavirus Is Airborne — but Health Advice Has Not Caught Up. *Nature* **2020**, *583* (7817), 510–513. https://doi.org/10.1038/d41586-020-02058-1.

(4) Mandavilli, A. 239 Experts With One Big Claim: The Coronavirus Is Airborne. *The New York Times*. July 4, 2020.

(5) WHO agrees with more than 200 medical experts that COVID-19 may spread via the air https://www.usatoday.com/in-depth/news/2020/04/03/coronavirus-protection-how-masks-might-stop-spread-through-coughs/5086553002/ (accessed Jul 26, 2020).

(6) Chao, C. Y. H.; Wan, M. P.; Morawska, L.; Johnson, G. R.; Ristovski, Z. D.; Hargreaves, M.; Mengersen, K.; Corbett, S.; Li, Y.; Xie, X.; Katoshevski, D. Characterization of Expiration Air Jets and Droplet Size Distributions Immediately at the Mouth Opening. *Journal of Aerosol Science* **2009**, *40* (2), 122–133. https://doi.org/10.1016/j.jaerosci.2008.10.003.

(7) Asadi, S.; Wexler, A. S.; Cappa, C. D.; Barreda, S.; Bouvier, N. M.; Ristenpart, W. D. Aerosol Emission and Superemission during Human Speech Increase with Voice Loudness. *Sci Rep* **2019**, *9* (1), 2348. https://doi.org/10.1038/s41598-019-38808-z.

(8) Xie, X.; Li, Y.; Sun, H.; Liu, L. Exhaled Droplets Due to Talking and Coughing. *J. R. Soc. Interface* **2009**, *6* (suppl_6). https://doi.org/10.1098/rsif.2009.0388.focus.

(9) Duguid, J. P. The Size and the Duration of Air-Carriage of Respiratory Droplets and Droplet-Nuclei. *Epidemiol. Infect.* **1946**, *44* (6), 471–479. https://doi.org/10.1017/S0022172400019288.

(10) Wei, J.; Li, Y. Airborne Spread of Infectious Agents in the Indoor Environment. *Am J Infect Control* **2016**, *44* (9 Suppl), S102-108. https://doi.org/10.1016/j.ajic.2016.06.003.

(11) Mittal, R.; Ni, R.; Seo, J.-H. The Flow Physics of COVID-19. *Journal of Fluid Mechanics* **2020**, *894*. https://doi.org/10.1017/jfm.2020.330.

(12) Petersen, E.; Koopmans, M.; Go, U.; Hamer, D. H.; Petrosillo, N.; Castelli, F.; Storgaard, M.; Khalili, S. A.; Simonsen, L. Comparing SARS-CoV-2 with SARS-CoV and Influenza Pandemics. *The Lancet Infectious Diseases* **2020**, *0* (0). https://doi.org/10.1016/S1473-3099(20)30484-9.

(13) van Doremalen, N.; Bushmaker, T.; Morris, D. H.; Holbrook, M. G.; Gamble, A.; Williamson, B. N.; Tamin, A.; Harcourt, J. L.; Thornburg, N. J.; Gerber, S. I.; Lloyd-Smith, J. O.; de Wit, E.; Munster, V. J. Aerosol and Surface Stability of SARS-CoV-2 as Compared with SARS-CoV-1. *N Engl J Med* **2020**, NEJMc2004973. https://doi.org/10.1056/NEJMc2004973.

(14) Coronavirus Disease (COVID-19) Situation Reports https://www.who.int/emergencies/diseases/novel-coronavirus-2019/situation-reports (accessed Jul 26, 2020).

(15) Morawska, L.; Tang, J. W.; Bahnfleth, W.; Bluyssen, P. M.; Boerstra, A.; Buonanno, G.; Cao, J.; Dancer, S.; Floto, A.; Querol, X.; Wierzbicka, A. How Can Airborne Transmission of COVID-19 Indoors Be Minimised? **2020**. https://doi.org/10.1016/j.envint.2020.105832.

(16) Cheng, V. C.-C.; Wong, S.-C.; Chuang, V. W.-M.; So, S. Y.-C.; Chen, J. H.-K.; Sridhar, S.; To, K. K.-W.; Chan, J. F.-W.; Hung, I. F.-N.; Ho, P.-L.; Yuen, K.-Y. The Role of Community-Wide Wearing of Face Mask for Control of Coronavirus Disease 2019 (COVID-19) Epidemic Due to SARS-CoV-2. *Journal of Infection* **2020**, *81* (1), 107–114. https://doi.org/10.1016/j.jinf.2020.04.024.

(17) Eikenberry, S. E.; Mancuso, M.; Iboi, E.; Phan, T.; Eikenberry, K.; Kuang, Y.; Kostelich, E.; Gumel, A. B. To Mask or Not to Mask: Modeling the Potential for Face Mask Use by the General Public to Curtail the COVID-19 Pandemic. *Infectious Disease Modelling* **2020**, *5*, 293–308. https://doi.org/10.1016/j.idm.2020.04.001.

(18) COVID-19 Map https://coronavirus.jhu.edu/map.html (accessed Aug 2, 2020).

(19) Ćirković, M. M. The Temporal Aspect of the Drake Equation and SETI. *Astrobiology* **2004**, *4* (2), 225–231. https://doi.org/10.1089/153110704323175160.

(20) Burchell, M. J. W(h)Ither the Drake Equation? *International Journal of Astrobiology; Cambridge* **2006**, *5* (3), 243–250.

(21) Buonanno, G.; Stabile, L.; Morawska, L. Estimation of Airborne Viral Emission: Quanta Emission Rate of SARS-CoV-2 for Infection Risk Assessment. *Environment International* **2020**, *141*, 105794. https://doi.org/10.1016/j.envint.2020.105794.

(22) Yezli, S.; Otter, J. A. Minimum Infective Dose of the Major Human Respiratory and Enteric Viruses Transmitted Through Food and the Environment. *Food Environ Virol* **2011**, *3* (1), 1–30. https://doi.org/10.1007/s12560-011-9056-7.

(23) Gammaitoni, L.; Nucci, M. C. Using a Mathematical Model to Evaluate the Efficacy of TB Control Measures. *Emerg Infect Dis* **1997**, *3* (3), 335–342.

(24) Riley, E. C.; Murphy, G.; Riley, R. L. Airborne Spread of Measles in a Suburban Elementary School. *Am. J. Epidemiol.* **1978**, *107* (5), 421–432. https://doi.org/10.1093/oxfordjournals.aje.a112560.

(25) Johnson, G. R.; Morawska, L. The Mechanism of Breath Aerosol Formation. *Journal of Aerosol Medicine and Pulmonary Drug Delivery* **2009**, *22* (3), 229–237. https://doi.org/10.1089/jamp.2008.0720.

(26) Lee, J.; Yoo, D.; Ryu, S.; Ham, S.; Lee, K.; Yeo, M.; Min, K.; Yoon, C. Quantity, Size Distribution, and Characteristics of Cough-Generated Aerosol Produced by Patients with an Upper Respiratory Tract Infection. *Aerosol Air Qual. Res.* **2019**, *19* (4), 840–853. https://doi.org/10.4209/aaqr.2018.01.0031.

(27) Fennelly, K. P. Particle Sizes of Infectious Aerosols: Implications for Infection Control. *The Lancet Respiratory Medicine* **2020**, *0* (0). https://doi.org/10.1016/S2213-2600(20)30323-4.

(28) Milton, D. K.; Fabian, M. P.; Cowling, B. J.; Grantham, M. L.; McDevitt, J. J. Influenza Virus Aerosols in Human Exhaled Breath: Particle Size, Culturability, and Effect of Surgical Masks. *PLoS Pathog* **2013**, *9* (3), e1003205. https://doi.org/10.1371/journal.ppat.1003205.

(29) Lindsley, W. G.; Pearce, T. A.; Hudnall, J. B.; Davis, K. A.; Davis, S. M.; Fisher, M. A.; Khakoo, R.; Palmer, J. E.; Clark, K. E.; Celik, I.; Coffey, C. C.; Blachere, F. M.; Beezhold, D. H. Quantity and Size Distribution of Cough-Generated Aerosol Particles Produced by Influenza Patients During and After Illness. *J Occup Environ Hyg* **2012**, *9* (7), 443–449. https://doi.org/10.1080/15459624.2012.684582.

(30) Stadnytskyi, V.; Bax, C. E.; Bax, A.; Anfinrud, P. The Airborne Lifetime of Small Speech Droplets and Their Potential Importance in SARS-CoV-2 Transmission. *PNAS* **2020**, *117* (22), 11875–11877. https://doi.org/10.1073/pnas.2006874117.

(31) Han, Z. Y.; Weng, W. G.; Huang, Q. Y. Characterizations of Particle Size Distribution of the Droplets Exhaled by Sneeze. *J. R. Soc. Interface* **2013**, *10* (88), 20130560. https://doi.org/10.1098/rsif.2013.0560.

(32) Bai, Y.; Yao, L.; Wei, T.; Tian, F.; Jin, D.-Y.; Chen, L.; Wang, M. Presumed Asymptomatic Carrier Transmission of COVID-19. *JAMA* **2020**, *323* (14), 1406–1407. https://doi.org/10.1001/jama.2020.2565.

(33) Nicolò, A.; Girardi, M.; Bazzucchi, I.; Felici, F.; Sacchetti, M. Respiratory Frequency and Tidal Volume during Exercise: Differential Control and Unbalanced Interdependence. *Physiol Rep* **2018**, *6* (21). https://doi.org/10.14814/phy2.13908.

(34) Morawska, L.; Johnson, G. R.; Ristovski, Z. D.; Hargreaves, M.; Mengersen, K.; Corbett, S.; Chao, C. Y. H.; Li, Y.; Katoshevski, D. Size Distribution and Sites of Origin of Droplets Expelled from the Human Respiratory Tract during Expiratory Activities. *Journal of Aerosol Science* **2009**, *40* (3), 256–269. https://doi.org/10.1016/j.jaerosci.2008.11.002.

(35) Edwards, D. A.; Man, J. C.; Brand, P.; Katstra, J. P.; Sommerer, K.; Stone, H. A.; Nardell, E.; Scheuch, G. Inhaling to Mitigate Exhaled Bioaerosols. *Proceedings of the National Academy of Sciences* **2004**, *101* (50), 17383–17388. https://doi.org/10.1073/pnas.0408159101.

(36) Johnson, G. R.; Morawska, L.; Ristovski, Z. D.; Hargreaves, M.; Mengersen, K.; Chao, C. Y. H.; Wan, M. P.; Li, Y.; Xie, X.; Katoshevski, D.; Corbett, S. Modality of Human Expired Aerosol Size Distributions. *Journal of Aerosol Science* **2011**, *42* (12), 839–851. https://doi.org/10.1016/j.jaerosci.2011.07.009.

(37) Fiegel, J.; Clarke, R.; Edwards, D. A. Airborne Infectious Disease and the Suppression of Pulmonary Bioaerosols. *Drug Discovery Today* **2006**, *11* (1–2), 51–57. https://doi.org/10.1016/S1359-6446(05)03687-1.

(38) Xie, X.; Li, Y.; Chwang, A. T. Y.; Ho, P. L.; Seto, W. H. How Far Droplets Can Move in Indoor Environments--Revisiting the Wells Evaporation-Falling Curve. *Indoor Air* **2007**, *17* (3), 211–225. https://doi.org/10.1111/j.1600-0668.2007.00469.x.

(39) Lohse, D. Die Abstandsregel in Zeiten von Corona. *Physik Journal* **2020**, *19* (5), 18–19.

(40) Gralton, J.; Tovey, E.; McLaws, M.-L.; Rawlinson, W. D. The Role of Particle Size in Aerosolised Pathogen Transmission: A Review. *Journal of Infection* **2011**, *62* (1), 1–13. https://doi.org/10.1016/j.jinf.2010.11.010.

(41) Small, M.; Tse, C. K.; Walker, D. M. Super-Spreaders and the Rate of Transmission of the SARS Virus. *Physica D* **2006**, *215* (2), 146–158. https://doi.org/10.1016/j.physd.2006.01.021.

(42) Wells, W. F. ON AIR-BORNE INFECTION: STUDY II. DROPLETS AND DROPLET NUCLEI. *American journal of Epidemiology* **1934**, *20* (3), 611–618.

(43) Dbouk, T.; Drikakis, D. On Coughing and Airborne Droplet Transmission to Humans. *Physics of Fluids* **2020**, *32* (5), 053310. https://doi.org/10.1063/5.0011960.

(44) Drivas, P. J.; Valberg, P. A.; Murphy, B. L.; Wilson, R. Modeling Indoor Air Exposure from Short-Term Point Source Releases. *Indoor Air* **1996**, *6* (4), 271–277. https://doi.org/10.1111/j.1600-0668.1996.00006.x.

(45) Licina, D.; Melikov, A.; Pantelic, J.; Sekhar, C.; Tham, K. W. Human Convection Flow in Spaces with and without Ventilation: Personal Exposure to Floor-Released Particles and Cough-Released Droplets. *Indoor Air* **2015**, *25* (6), 672–682. https://doi.org/10.1111/ina.12177.

(46) Salmanzadeh, M.; Zahedi, Gh.; Ahmadi, G.; Marr, D. R.; Glauser, M. Computational Modeling of Effects of Thermal Plume Adjacent to the Body on the Indoor Airflow and Particle Transport. *Journal of Aerosol Science* **2012**, *53*, 29–39. https://doi.org/10.1016/j.jaerosci.2012.05.005.

(47) Craven, B. A.; Settles, G. S. A Computational and Experimental Investigation of the Human Thermal Plume. *J. Fluids Eng* **2006**, *128* (6), 1251–1258. https://doi.org/10.1115/1.2353274.

(48) Bhagat, R. K.; Wykes, M. S. D.; Dalziel, S. B.; Linden, P. F. Effects of Ventilation on the Indoor Spread of COVID-19. *Journal of Fluid Mechanics to Appear.*

(49) Roth, M. Review of Atmospheric Turbulence over Cities. *Quarterly Journal of the Royal Meteorological Society* **2000**, *126* (564), 941–990. https://doi.org/10.1002/qj.49712656409.

(50) Sutton, O. G. *Atmospheric Turbulence*; Routledge, 2020.

(51) Liu, L.; Li, Y.; Nielsen, P. V.; Wei, J.; Jensen, R. L. Short-Range Airborne Transmission of Expiratory Droplets between Two People. *Indoor Air* **2017**, *27* (2), 452–462. https://doi.org/10.1111/ina.12314.

(52) Kwon, S.-B.; Park, J.; Jang, J.; Cho, Y.; Park, D.-S.; Kim, C.; Bae, G.-N.; Jang, A. Study on the Initial Velocity Distribution of Exhaled Air from Coughing and Speaking. *Chemosphere* **2012**, *87* (11), 1260–1264. https://doi.org/10.1016/j.chemosphere.2012.01.032.

(53) Wei, J.; Li, Y. Human Cough as a Two-Stage Jet and Its Role in Particle Transport. *PLoS One* **2017**, *12* (1). https://doi.org/10.1371/journal.pone.0169235.

(54) Abkarian, M.; Mendez, S.; Xue, N.; Yang, F.; Stone, H. A. Speech Can Produce Jet-like Transport Relevant to Asymptomatic Spreading of Virus. *arXiv:2006.10671 [physics]* **2020**.

(55) Pyankov, O. V.; Bodnev, S. A.; Pyankova, O. G.; Agranovski, I. E. Survival of Aerosolized Coronavirus in the Ambient Air. *Journal of Aerosol Science* **2018**, *115*, 158–163. https://doi.org/10.1016/j.jaerosci.2017.09.009.

(56) Schuit, M.; Ratnesar-Shumate, S.; Yolitz, J.; Williams, G.; Weaver, W.; Green, B.; Miller, D.; Krause, M.; Beck, K.; Wood, S.; Holland, B.; Bohannon, J.; Freeburger, D.; Hooper, I.; Biryukov, J.; Altamura, L. A.; Wahl, V.; Hevey, M.; Dabisch, P. Airborne SARS-CoV-2 Is Rapidly Inactivated by Simulated Sunlight. *J. Infect. Dis.* **2020**, *222* (4), 564–571. https://doi.org/10.1093/infdis/jiaa334.

(57) Evidence for Ultraviolet Radiation Decreasing COVID-19 Growth Rates: Global Estimates and Seasonal Implications.

(58) Yang, W.; Elankumaran, S.; Marr, L. C. Relationship between Humidity and Influenza A Viability in Droplets and Implications for Influenza's Seasonality. *PLoS One* **2012**, *7* (10). https://doi.org/10.1371/journal.pone.0046789.

(59) The change in seasons may or may not affect the spread of COVID-19 | National Academies http://sites.nationalacademies.org/BasedOnScience/change-in-seasons-may-or-may-not-affect-the-spread-of-COVID-19/index.htm (accessed Jul 31, 2020).

(60) Hallett, S.; Toro, F.; Ashurst, J. V. Physiology, Tidal Volume. In *StatPearls*; StatPearls Publishing: Treasure Island (FL), 2020.

(61) Santuz, P.; Baraldi, E.; Filippone, M.; Zacchello, F. Exercise Performance in Children with Asthma: Is It Different from That of Healthy Controls? *European Respiratory Journal* **1997**, *10* (6), 1254–1260.

(62) Aitken, M. L.; Franklin, J. L.; Pierson, D. J.; Schoene, R. B. Influence of Body Size and Gender on Control of Ventilation. *Journal of Applied Physiology* **1986**, *60* (6), 1894–1899. https://doi.org/10.1152/jappl.1986.60.6.1894.

(63) Radford, E. P. Ventilation Standards for Use in Artificial Respiration. *Journal of Applied Physiology* **1955**, *7* (4), 451–460. https://doi.org/10.1152/jappl.1955.7.4.451.

(64) Lee, P.-I.; Hu, Y.-L.; Chen, P.-Y.; Huang, Y.-C.; Hsueh, P.-R. Are Children Less Susceptible to COVID-19? *J Microbiol Immunol Infect* **2020**, *53* (3), 371–372. https://doi.org/10.1016/j.jmii.2020.02.011.

(65) Kass, D. A.; Duggal, P.; Cingolani, O. Obesity Could Shift Severe COVID-19 Disease to Younger Ages. *The Lancet* **2020**, *395* (10236), 1544–1545. https://doi.org/10.1016/S0140-6736(20)31024-2.

(66) Nikitin, N.; Petrova, E.; Trifonova, E.; Karpova, O. Influenza Virus Aerosols in the Air and Their Infectiousness. *Adv Virol* **2014**, *2014*, 859090. https://doi.org/10.1155/2014/859090.

(67) Tao, X.; Garron, T.; Agrawal, A. S.; Algaissi, A.; Peng, B.-H.; Wakamiya, M.; Chan, T.-S.; Lu, L.; Du, L.; Jiang, S.; Couch, R. B.; Tseng, C.-T. K. Characterization and Demonstration of the Value of a Lethal Mouse Model of Middle East Respiratory Syndrome Coronavirus Infection and Disease. *Journal of Virology* **2016**, *90* (1), 57–67. https://doi.org/10.1128/JVI.02009-15.

(68) Alford, R. H.; Kasel, J. A.; Gerone, P. J.; Knight, V. Human Influenza Resulting from Aerosol Inhalation. *Proceedings of the Society for Experimental Biology and Medicine* **1966**, *122* (3), 800–804. https://doi.org/10.3181/00379727-122-31255.

(69) Nicas, M.; Nazaroff, W. W.; Hubbard, A. Toward Understanding the Risk of Secondary Airborne Infection: Emission of Respirable Pathogens. *J Occup Environ Hyg* **2010**, *2* (3), 143–154. https://doi.org/10.1080/15459620590918466.

(70) Konda, A.; Prakash, A.; Moss, G. A.; Schmoldt, M.; Grant, G. D.; Guha, S. Aerosol Filtration Efficiency of Common Fabrics Used in Respiratory Cloth Masks. *ACS Nano* **2020**. https://doi.org/10.1021/acsnano.0c03252.

(71) Zangmeister, C. D.; Radney, J. G.; Vicenzi, E. P.; Weaver, J. L. Filtration Efficiencies of Nanoscale Aerosol by Cloth Mask Materials Used to Slow the Spread of SARS-CoV-2. *ACS Nano* **2020**, *14* (7), 9188–9200. https://doi.org/10.1021/acsnano.0c05025.

(72) Leung, N. H. L.; Chu, D. K. W.; Shiu, E. Y. C.; Chan, K.-H.; McDevitt, J. J.; Hau, B. J. P.; Yen, H.-L.; Li, Y.; Ip, D. K. M.; Peiris, J. S. M.; Seto, W.-H.; Leung, G. M.; Milton, D. K.; Cowling, B. J. Respiratory Virus Shedding in Exhaled Breath and Efficacy of Face Masks. *Nature Medicine* **2020**, *26* (5), 676–680. https://doi.org/10.1038/s41591-020-0843-2.

(73) Verma, S.; Dhanak, M.; Frankenfield, J. Visualizing the Effectiveness of Face Masks in Obstructing Respiratory Jets. *Physics of Fluids* **2020**, *32* (6), 061708. https://doi.org/10.1063/5.0016018.

(74) Dbouk, T.; Drikakis, D. On Respiratory Droplets and Face Masks. *Phys Fluids* **2020**, *32* (6), 063303. https://doi.org/10.1063/5.0015044.

(75) Pendar, M.-R.; Páscoa, J. C. Numerical Modeling of the Distribution of Virus Carrying Saliva Droplets during Sneeze and Cough. *Physics of Fluids* **2020**, *32* (8), 083305. https://doi.org/10.1063/5.0018432.

(76) Tang, J. W.; Liebner, T. J.; Craven, B. A.; Settles, G. S. A Schlieren Optical Study of the Human Cough with and without Wearing Masks for Aerosol Infection Control. *J. R. Soc. Interface* **2009**, *6* (suppl_6). https://doi.org/10.1098/rsif.2009.0295.focus.

(77) Bourouiba, L.; Dehandschoewercker, E.; Bush, J. W. M. Violent Expiratory Events: On Coughing and Sneezing. *J. Fluid Mech.* **2014**, *745*, 537–563. https://doi.org/10.1017/jfm.2014.88.

(78) Fischer, E. P.; Fischer, M. C.; Grass, D.; Henrion, I.; Warren, W. S.; Westman, E. Low-Cost Measurement of Facemask Efficacy for Filtering Expelled Droplets during Speech. *Science Advances* **2020**, eabd3083. https://doi.org/10.1126/sciadv.abd3083.

(79) Chan, J. F.-W.; Yuan, S.; Zhang, A. J.; Poon, V. K.-M.; Chan, C. C.-S.; Lee, A. C.-Y.; Fan, Z.; Li, C.; Liang, R.; Cao, J.; Tang, K.; Luo, C.; Cheng, V. C.-C.; Cai, J.-P.; Chu, H.; Chan, K.-H.; To, K. K.-W.; Sridhar, S.;

Yuen, K.-Y. Surgical Mask Partition Reduces the Risk of Noncontact Transmission in a Golden Syrian Hamster Model for Coronavirus Disease 2019 (COVID-19). *Clin Infect Dis*. https://doi.org/10.1093/cid/ciaa644.

(80) Bird, R. B.; Stewart, W. E.; Lightfoot, E. N. *Transport Phenomena*, 2nd ed.; Wiley, 2006.

(81) Tennekes, H.; Lumley, J. L. *A First Course in Turbulence*; MIT Press, 2018.

(82) Zarruk, G. A.; Cowen, E. A. Simultaneous Velocity and Passive Scalar Concentration Measurements in Low Reynolds Number Neutrally Buoyant Turbulent Round Jets. *Exp Fluids* **2008**, *44* (6), 865–872. https://doi.org/10.1007/s00348-007-0441-9.

(83) Yang, A.A.Pahlavan, S. Mendez, M. Abkarian, and H.A. Stone, ", F.; Pahlavan, A. A.; Mendez, S.; Abkarian, M.; Stone, H. Towards Improved Social Distancing Guidelines: Space and Time Dependence of Virus Transmission from Speech-Driven Aerosol Transport Between Two Individuals. **2020**, No. Preprint.

(84) Tang, J. W.; Nicolle, A. D.; Klettner, C. A.; Pantelic, J.; Wang, L.; Suhaimi, A. B.; Tan, A. Y. L.; Ong, G. W. X.; Su, R.; Sekhar, C.; Cheong, D. D. W.; Tham, K. W. Airflow Dynamics of Human Jets: Sneezing and Breathing - Potential Sources of Infectious Aerosols. *PLOS ONE* **2013**, *8* (4), e59970. https://doi.org/10.1371/journal.pone.0059970.

(85) Sykes, R. I.; Henn, D. S. Large-Eddy Simulation of Concentration Fluctuations in a Dispersing Plume. *Atmospheric Environment. Part A. General Topics* **1992**, *26* (17), 3127–3144. https://doi.org/10.1016/0960-1686(92)90470-6.

(86) Vinkovic, I.; Aguirre, C.; Simons, S. Large-Eddy Simulation and Lagrangian Stochastic Modeling of Passive Scalar Dispersion in a Turbulent Boundary Layer. *Journal of Turbulence* **2006**, *7*, N30. https://doi.org/10.1080/14685240600595537.

(87) Wu, W.; Piomelli, U. Reynolds-Averaged and Wall-Modelled Large-Eddy Simulations of Impinging Jets with Embedded Azimuthal Vortices. *European Journal of Mechanics - B/Fluids* **2016**, *55*, 348–359. https://doi.org/10.1016/j.euromechflu.2015.06.008.

(88) Norton, K.; Norton, L.; Sadgrove, D. Position Statement on Physical Activity and Exercise Intensity Terminology. *Journal of Science and Medicine in Sport* **2010**, *13* (5), 496–502. https://doi.org/10.1016/j.jsams.2009.09.008.

(89) Valli, G.; Internullo, M.; Ferrazza, A. M.; Onorati, P.; Cogo, A.; Palange, P. Minute Ventilation and Heart Rate Relationship for Estimation of the Ventilatory Compensation Point at High Altitude: A Pilot Study. *Extrem Physiol Med* **2013**, *2*, 7. https://doi.org/10.1186/2046-7648-2-7.

(90) Sze To, G. N.; Chao, C. Y. H. Review and Comparison between the Wells–Riley and Dose-response Approaches to Risk Assessment of Infectious Respiratory Diseases. *Indoor Air* **2010**, *20* (1), 2–16. https://doi.org/10.1111/j.1600-0668.2009.00621.x.

(91) Sze To, G. N.; Wan, M. P.; Chao, C. Y. H.; Wei, F.; Yu, S. C. T.; Kwan, J. K. C. A Methodology for Estimating Airborne Virus Exposures in Indoor Environments Using the Spatial Distribution of Expiratory Aerosols and Virus Viability Characteristics. *Indoor Air* **2008**, *18* (5), 425–438. https://doi.org/10.1111/j.1600-0668.2008.00544.x.

(92) Gandhi, M.; Beyrer, C.; Goosby, E. Masks Do More Than Protect Others During COVID-19: Reducing the Inoculum of SARS-CoV-2 to Protect the Wearer. *J Gen Intern Med* **2020**. https://doi.org/10.1007/s11606-020-06067-8.

(93) *An Introduction to Probability Theory and Its Applications*; Wiley; Vol. 2.

(94) Noakes, C. J.; Beggs, C. B.; Sleigh, P. A.; Kerr, K. G. Modelling the Transmission of Airborne Infections in Enclosed Spaces. *Epidemiol Infect* **2006**, *134* (5), 1082–1091. https://doi.org/10.1017/S0950268806005875.

(95) Stutt, R. O. J. H.; Retkute, R.; Bradley, M.; Gilligan, C. A.; Colvin, J. A Modelling Framework to Assess the Likely Effectiveness of Facemasks in Combination with 'Lock-down' in Managing the COVID-19 Pandemic. *Proc Math Phys Eng Sci* **2020**, *476* (2238). https://doi.org/10.1098/rspa.2020.0376.

(96) Watanabe, T.; Bartrand, T. A.; Weir, M. H.; Omura, T.; Haas, C. N. Development of a Dose-Response Model for SARS Coronavirus. *Risk Anal* **2010**, *30* (7), 1129–1138. https://doi.org/10.1111/j.1539-6924.2010.01427.x.

(97) Ferry, J.; Balachandar, S. A Fast Eulerian Method for Disperse Two-Phase Flow. *International Journal of Multiphase Flow 27* (7), 1199–1226.

(98) Aiyer, A. K.; Yang, D.; Chamecki, M.; Meneveau, C. A Population Balance Model for Large Eddy Simulation of Polydisperse Droplet Evolution. *Journal of Fluid Mechanics* **2019**, *878*, 700–739. https://doi.org/10.1017/jfm.2019.649.

(99) Yang, D.; Chen, B.; Socolofsky, S. A.; Chamecki, M.; Meneveau, C. Large-Eddy Simulation and Parameterization of Buoyant Plume Dynamics in Stratified Flow. *Journal of Fluid Mechanics* **2016**, *794*, 798–833. https://doi.org/10.1017/jfm.2016.191.

(100) Germano, M.; Piomelli, U.; Moin, P.; Cabot, W. H. A Dynamic Subgrid-scale Eddy Viscosity Model. *Physics of Fluids A: Fluid Dynamics* **1991**, *3* (7), 1760–1765. https://doi.org/10.1063/1.857955.

(101) Meneveau, C.; Lund, T. S.; Cabot, W. H. A Lagrangian Dynamic Subgrid-Scale Model of Turbulence. *Journal of Fluid Mechanics* **1996**, *319*, 353–385. https://doi.org/10.1017/S0022112096007379.

(102) Scalo, C.; Boegman, L.; Piomelli, U. Large-Eddy Simulation and Low-Order Modeling of Sediment-Oxygen Uptake in a Transitional Oscillatory Flow. *Journal of Geophysical Research: Oceans* **2013**, *118* (4), 1926–1939. https://doi.org/10.1002/jgrc.20113.

| No. | Axis Legend | Name in Paper (ID, structure, yarns/inch) | $\overline{FE}$ Lower Bound | $\overline{FE}$ Upper Bound | No. | Axis Legend | Name in Paper (ID, structure, yarns/inch) | $\overline{FE}$ Lower Bound | $\overline{FE}$ Upper Bound |
|---|---|---|---|---|---|---|---|---|---|
| 1 | Cotton 1 | Cotton 1, plain weave, 152 | 26.0 | 81.9 | 18 | Polyester 2 | Polyester 2, plain weave, 152 | 34.6 | 85.1 |
| 2 | Cotton 2 | Cotton 2, plain weave, 152 | 12.0 | 72.8 | 19 | Polyester 3 | Polyester 3, poplin weave, 152 | 30.6 | 82.3 |
| 3 | Cotton 3 | Cotton 3, plain weave, 152 | 24.1 | 79.8 | 20 | Polyester 4 | Polyester 4, poplin weave, 152 | 45.0 | 87.7 |
| 4 | Cotton 4 | Cotton 4, plain weave, 229 | 66.4 | 91.8 | 21 | Polyester 5 | Polyester 5, poplin weave, 229 | 28.0 | 79.0 |
| 5 | Cotton 5 | Cotton 5, sateen, weave, 356 | 22.1 | 77.7 | 22 | Syn Blend 1 | Syn Blend 1, poplin weave, 330 | 20.0 | 78.7 |
| 6 | Cotton 6 | Cotton 6, satin weave, 812 | 38.9 | 86.0 | 23 | Syn Blend 2 | Syn Blend 2, twill weave, 203 | 60.7 | 90.6 |
| 7 | Cotton 7 | Cotton 7, poplin weave, 89 | 18.5 | 71.8 | 24 | Syn Blend 3 | Syn Blend 3, waffle weave, 25 | 24.6 | 67.9 |
| 8 | Cotton 8 | Cotton 8, block weave, 102 | 37.7 | 80.1 | 25 | Syn Blend 4 | Syn Blend 4, poplin weave, 229 | 27.7 | 80.9 |
| 9 | Cotton 9 | Cotton 9, poplin weave, 152 | 30.1 | 81.3 | 26 | Poly/Cott blend 1 | Poly/Cott blend 1, plain weave, 229 | 15.6 | 76.1 |
| 10 | Cotton 10 | Cotton 10, poplin weave, 152 | 38.4 | 85.3 | 27 | Poly/Cott blend 2 | Poly/Cott blend 2, poplin weave, 152 | 32.6 | 84.5 |
| 11 | Cotton 11 | Cotton 11, plain weave, 152 | 4.4 | 63.2 | 28 | Poly/Cott blend 3 | Poly/Cott blend 3, twill weave, 229 | 39.0 | 86.9 |
| 12 | Cotton 12 | Cotton 12, knit, 100 | 20.7 | 76.8 | 29 | Poly/Cott blend 4 | Poly/Cott blend 4, twill weave, 203 | 18.6 | 78.3 |
| 13 | Cotton 13 | Cotton 13, satin weave, 600 | 35.8 | 85.3 | 30 | 1 x Cotton 10, 1 x Polyester 4 | 1 x Cotton 10, 1 x Polyester 4 | 41.1 | 86.6 |
| 14 | Cotton 14 | Cotton 14, Quilt Batting | 57.9 | 90.8 | 31 | 1 x Cotton 13, 2 x Syn Blend 4 | 1 x Cotton 13, 2 x Syn Blend 4 | 26.9 | 82.0 |
| 15 | Wool | Wool, plain weave, 101 | 16.1 | 71.3 | 32 | 1 x Cotton 10, 1 x Polyester 6 | 1 x Cotton 10, 1 x Polyester 6 | 43.8 | 87.2 |
| 16 | Nylon | Nylon, poplin weave, 127 | 21.4 | 81.6 | 33 | 1 x Cotton 13, 1 x Poly/Cotton 4 | 1 x Cotton 13, 1 x Poly/Cotton 4 | 27.9 | 79.3 |
| 17 | Rayon | Rayon, plain weave, 152 | 4.6 | 54.3 | 34 | Surgical Mask 1 | Surgical Mask 1, multi-layer | 49.3 | 88.8 |

*Table 1. Description of the various fabrics/samples from Zangmeister et al.[71] that are included in the current study.*